\documentclass{sigchi-ext}
\usepackage[T1]{fontenc}
\usepackage{textcomp}
\usepackage[scaled=.92]{helvet} 
\usepackage{graphicx} 
\usepackage{balance}  
\usepackage{booktabs} 
\usepackage{ccicons}  
\usepackage{ragged2e} 

\def\plaintitle{SIGCHI Extended Abstracts Sample File: Note Initial
  Caps} 
\def\emptyauthor{}
\def\plainkeywords{eXplainable AI; Technological Comfort; Personality Traits}

\title{The Drawback of Insight: Detailed Explanations Can Reduce Agreement with XAI}

\numberofauthors{6}
\author{%
  \alignauthor{%
    \textbf{Sabid Bin Habib Pias$^*$}\\
    \affaddr{Indiana University} \\
    \affaddr{Indiana, USA} \\
     }\alignauthor{%
    \textbf{Taslima Akter}\\
    \affaddr{UC Irvine}\\
    \affaddr{California, USA}\\
     } \vfil \alignauthor{%
    \textbf{Alicia Freel}\\
    \affaddr{Indiana University} \\
    \affaddr{Indiana, USA} \\
     }\alignauthor{%
    \textbf{Donald Williamson}\\
    \affaddr{Ohio State University}\\
    \affaddr{Ohio, USA}\\
     } \vfil \alignauthor{%
    \textbf{Timothy Trammel}\\   
    \affaddr{UC Davis}\\
    \affaddr{California, USA}\\
    }\alignauthor{%
    \textbf{Apu Kapadia}\\
    \affaddr{Indiana University} \\
    \affaddr{Indiana, USA} \\
     } }

\definecolor{linkColor}{RGB}{6,125,233}
\hypersetup{%
  pdftitle={\plaintitle},
  pdfauthor={\emptyauthor},
  pdfkeywords={\plainkeywords},
  bookmarksnumbered,
  pdfstartview={FitH},
  colorlinks,
  citecolor=black,
  filecolor=black,
  linkcolor=black,
  urlcolor=linkColor,
  breaklinks=true,
}

\begin{document}

\CopyrightYear{2020}
\setcopyright{rightsretained}
\conferenceinfo{ACM CHI Workshop on Human-Centered Explainable AI}{May 12, 2024, Honolulu, HI, USA}
\isbn{978-1-4503-6819-3/20/04}
\doi{https://doi.org/10.1145/3334480.XXXXXXX}
\copyrightinfo{\acmcopyright}

\maketitle

\RaggedRight{} 
\begin{abstract}
With the emergence of Artificial Intelligence (AI)-based decision-making, explanations help increase new technology adoption through enhanced trust and reliability. However, our experimental study challenges the notion that every user universally values explanations. We argue that the agreement with AI suggestions, whether accompanied by explanations or not, is influenced by individual differences in personality traits and the users' comfort with technology. We found that people with higher neuroticism and lower technological comfort showed more agreement with the recommendations without explanations. As more users become exposed to eXplainable AI (XAI) and AI-based systems, we argue that the XAI design should not provide explanations for users with high neuroticism and low technology comfort. Prioritizing user personalities in XAI systems will help users become better collaborators of AI systems. 
\end{abstract}





\section{Introduction}

As an attempt to make eXplainable AI (XAI) more understandable and user-friendly, studies have explored the efficacy of single-style text explanation in varying AI-based solutions~\cite{kouki2019personalized, voorhees_system_2021}. Research showed that users have a higher perception of trust and transparency with text-based explanations with adequate information ~\cite{ooge2022explaining, eiband_impact_2019}. Moreover, people's acceptance of new information from either a human or an AI agent depends on their personality traits~\cite{park_who_2022, medders_personality_2017, kaya_roles_2024}. However, the acceptance of explanations by XAI-based systems based on the relationship between individual personality traits and the type of explanation presented to the user is understudied. Therefore, we seek to understand how users with different personalities accept explanations containing varying levels of information in an AI classifier system. In our study, we investigated the influence of three types of single-style text explanations: 1) no explanation, 2) placebic, and 3) meaningful explanations~\cite{eiband_impact_2019} based on participants' personality traits (neuroticism and conscientiousness) and technology comfort.
\marginpar{%
  \vspace{-220pt} \fbox{%
    \begin{minipage}{0.995\marginparwidth}
      \textbf{Personality traits considered in this study~\cite{barrick_big_1991}} \\
      \vspace{1pc} \textbf{Neuroticism:} the tendency to experience unpleasant emotions, such as anxiety, depression, and emotional volatility.  \\
      \vspace{1pc} \textbf{Conscientiousness:} the tendency to show self-discipline, aim for achievement, and having the attributes of organization, productivity, responsibility. \\
      \label{sidebar-personality}
    \end{minipage}} }

We designed an online user study (N = 224) with a simulated AI classifier interface to compare how participants with varying technological comfort, neuroticism, and conscientiousness agreed with the classifier with varying explanation types. We show that participants with low technology comfort and participants with higher neuroticism prefer prompts without any explanation, showing the need to evaluate user technology comfort and personality traits before deploying explanations in any XAI-based classifier or recommender. 
\section{Are Explanations Acceptable by Everyone?}

As XAI-based systems become popular in daily and productive tasks, more people are introduced to various AI technologies with explanations~\cite{rong_towards_2023}. However, the ultimate goal of explainable AI is to aid users, making it important to understand if the preference for explanations varies based on the unique personality traits of users. For this reason, we focus on participants' tendency to agree with different levels of information (no explanation, placebic, meaningful) based on their neuroticism, conscientiousness, and tech comfort. 

Neuroticism denotes the tendency or disposition to experience negative emotional states. People with high neuroticism are more likely to respond poorly to environmental stress and suspect negative outcomes from an ordinary situation~\cite{barrick_big_1991, widiger_neuroticism_2009}. Conscientiousness is about being careful, diligent, and thorough in daily activities. A person with high conscientiousness tends to pay attention to details and ensure that their deliverables are correct~\cite{roberts_what_2014}. Barnett et al. have shown that higher neuroticism negatively affects technology use and higher conscientiousness influences a person to use technologies~\cite{barnett_five-factor_2015}. Therefore, we contend that individuals exhibiting higher levels of neuroticism will opt for reduced information to mitigate distractions, and participants with higher conscientiousness will lean towards increased information provided by XAI-based systems to enhance task accuracy. Technological comfort refers to the ease and confidence of a person in using technological tools. This enables them to understand the technology and critically evaluate the information provided by the system. Park et al.~\cite{park_exploring_2019} posited that users' confidence in technology-provided information tends to be lower among individuals with limited technological competence than users with higher technological skills. Therefore, we hypothesize that participants with lower technological comfort will prefer minimal information in XAI-based systems. Based on previous findings that analyze the effect of neuroticism, conscientiousness, and tech comfort on user agreeability, ~\cite{roberts_what_2014, barnett_five-factor_2015}, we propose the following hypotheses in the context of XAI.\\
\textbf{H1:} Participants with lower tech comfort agree more with classifications \textit{without explanations} compared to classifications with \textit{meaningful explanations}  \\
\textbf{H2:} Participants with higher neuroticism agree more with classifications \textit{without explanations} compared to classifications with \textit{meaningful explanations} \\
\textbf{H3:} Participants with higher conscientiousness agree more with classifications with \textit{meaningful explanations} compared to classifications \textit{without explanations} \\

\section{Method}

\begin{marginfigure}[-75pt]
  \begin{minipage}{\marginparwidth}
    \centering
    \includegraphics[width=0.8\marginparwidth]{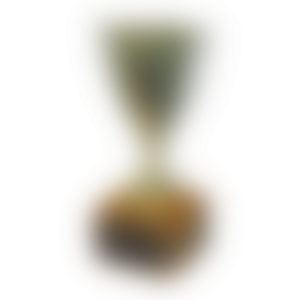}
    \caption{Sample blurred image of a trophy}
     \label{fig:trophy}
  \end{minipage}
\end{marginfigure}

\marginpar{%
  \vspace{30pt} 
  \fbox{%
    \begin{minipage}{\marginparwidth}
    \centering
      \textbf{Sample Explanations} \\
      \vspace{1pc}
      \textit{Correct Explanations} \\
      \RaggedRight
      \vspace{1pc} \textbf{No exp.:} This is a trophy.  \\
      \textbf{Placebic:} This is a trophy because it looks like a trophy. \\
      \textbf{Meaningful:} This is a trophy because it has a gold-colored cup and a white square base. \\
      
      \centering
      
      \vspace{1pc}
      \textit{Incorrect Explanations} \\
      \RaggedRight
      \vspace{1pc} \textbf{No explanation:} This is a medal.  \\
      \textbf{Placebic:} This is a medal because it looks like a medal. \\
      \textbf{Meaningful:} This is a medal because it is circular, gold, and hanging from a ribbon. 
    \end{minipage}
    } 
}

The study consisted of a norming study and an agreeability measurement study, both using a between-subject study design. In the norming study, 100 well-known object images were selected from the BOSS stimulus set ~\cite{brodeur2010bank}, and Gaussian blur filters with various levels were applied to determine the optimal blur level (sigma 115) that balanced identifiability. Participants (N = 60) viewed 50 blurred images randomly from a pool of 61 images. They identified the object and briefly described the characteristics that helped their identification. We used the most frequently stated characteristics to generate meaningful explanations and the second most frequently labeled objects as the incorrect labels for the agreement measure study. This ensured that the meaningful explanations for both correct and incorrect labels were based on object characteristics that humans found useful in identifying distorted images. In the agreeability measurement study, participants (N=225) were randomly assigned to one of three explanation conditions (no explanation, placebic, and meaningful) as part of the design between subjects. They were asked to rate their agreement with the AI's classification of the image on a 7-point Likert scale. Among the 60 images shown to each participant, 48 (80\%) were correctly labeled for the three explanation conditions. To ensure accurate balance, correct answers were counterbalanced using a modified Latin square design~\cite{rojas1957modified}. The study explored the relationship between personality, technology comfort, and participant agreement with the AI classifier for the three types of explanations. We used the Big 5 personality inventory (10-item version)  ~\cite{BeatriceMeasuring2007} to assess the neuroticism and conscientiousness of the participants via the mean score of the corresponding items. To measure technology comfort we simply asked participants how much they agreed/disagreed with the following statement ``Operating and using everyday technology is easy for me''. Additionally, we incorporated the technology acceptance questionnaire (TAM) ~\cite{davis1989technology}. The participants were recruited through Prolific~\footnote{\url{https://www.prolific.com/}} and were all adults from the United States. We paid \$7.00 to each participant for a completed study regardless of their response quality. The payment amount was calculated to be above \$15/hour which surpasses the recommended minimum wage~\cite{silberman2018responsible} in the study’s location. Statistical analysis used a linear mixed-effects model, treating explanation type, technology comfort, neuroticism, and conscientiousness as fixed effects, and participant and images as random effects.

\section{Findings}

\begin{figure}[!ht]
    
  \includegraphics[width=0.9\columnwidth]{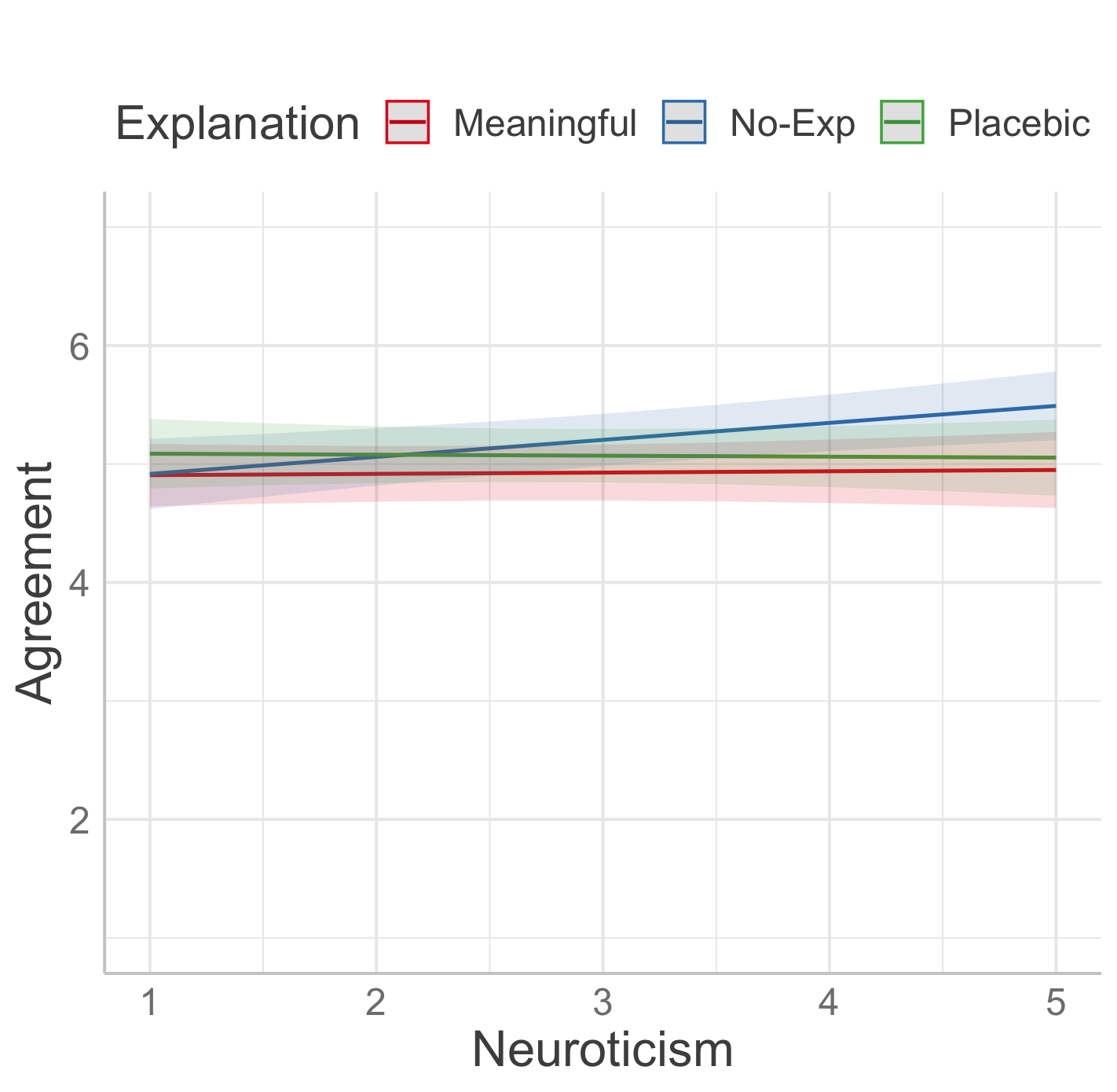}
  \caption{Relation of neuroticism and agreement to AI classifier}~\label{fig:neuro-exp}
 
\end{figure}

\begin{figure}[!ht]
    \centering
  \includegraphics[width=0.9\columnwidth]{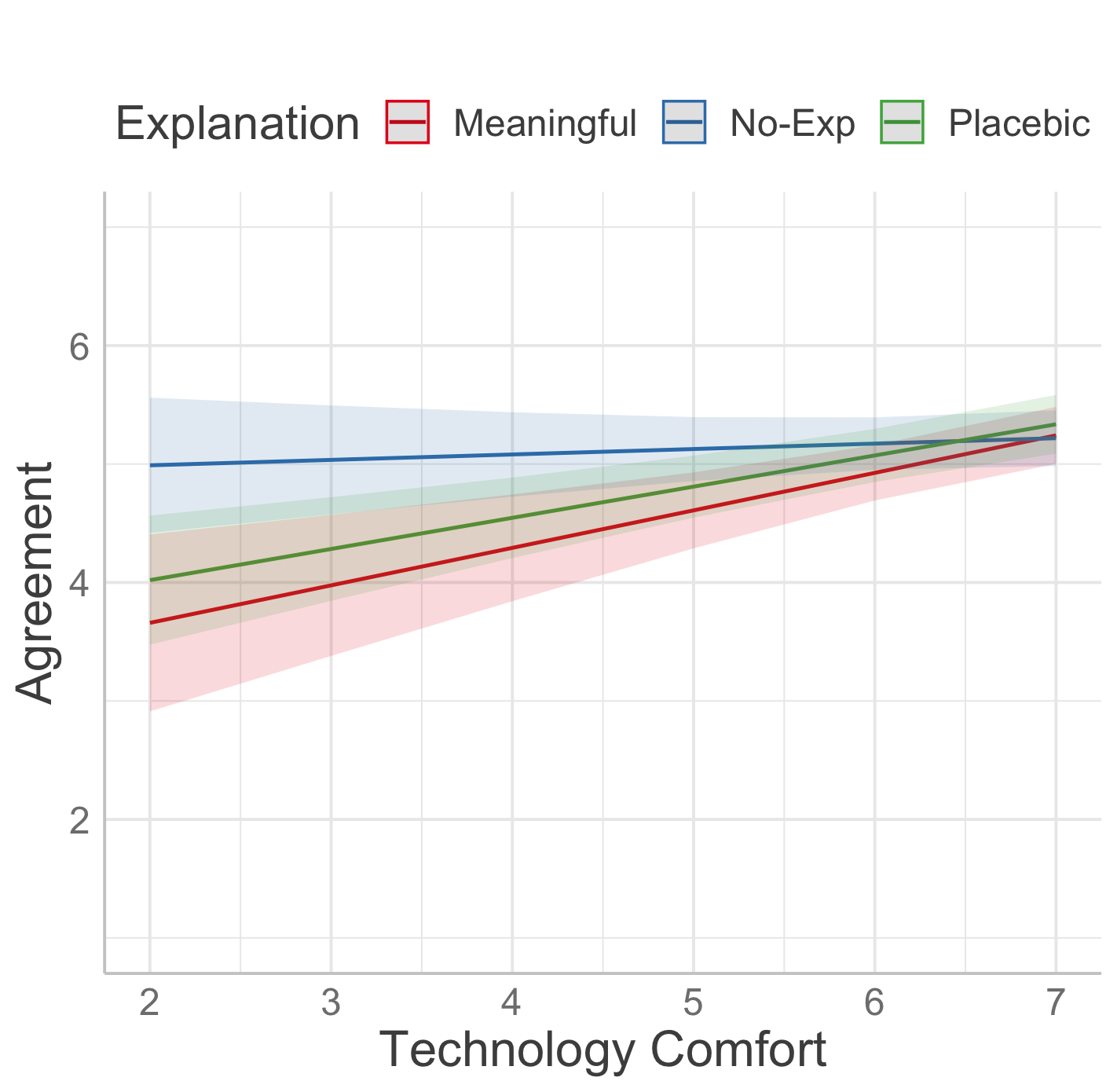}
  \caption{Relation of tech comfort and agreement to AI classifier}~\label{fig:tech-exp}
\end{figure}

\subsection{Effect of Technology Comfort and Explanation Type}

Participants indicating lower level of comfort with new technology were more likely to agree with the AI classifier when there were no explanations, compared to the other two conditions in which the AI classifier presented meaningful and placebic explanations (Fig. ~\ref{fig:tech-exp}). \emph{This finding suggests that users low in technological comfort are less comfortable navigating `new' technology that explains itself (supporting H1).}

\subsection{Effect of Neuroticism and Explanation Type}

From Table~\ref{tab:lmer-agree}, it is clear that individuals with greater neuroticism were more likely to agree with the recommendations that lacked explanations, compared to the recommendations with either meaningful or placebo information. In contrast, participants with lower levels of neuroticism showed no differences in their agreement with recommendations based on the type of explanation provided. (Fig. ~\ref{fig:neuro-exp}). \emph{The result indicates that participants with higher emotional volatility think that the classifier is more likely to be correct when the classifier provides less information compared to when the classifier provides more text in the form of meaningful and placebic explanations (supporting H2).}

\subsection{Effect of Conscientiousness and Explanation Type}

In contrast to our hypotheses, we did not find significant differences in participant agreement with the AI classifier presenting varying types of explanations (Table~\ref{tab:lmer-agree}). This result suggests that conscientiousness may not directly affect the reaction of the participant to the varying explanation types, failing to support H3.

\begin{table}
  \centering
  \begin{tabular}{l r r}
    
    {\small\textit{Effects}}
    & {\small \textit{Estimate}}
      & {\small \textit{Std. Error}} \\
    \midrule
    TC: Meaningful & $0.27^{**}$ & 0.1  \\
    TC: Placebic & $0.21^{*}$ & 0.08  \\
    Neuro: Meaningful & $-0.13^{*}$ & 0.06  \\
    Neuro: Placebic & $-0.15^{*}$ & 0.07  \\
    Conscience: Meaningful & $-0.15$ & 0.09  \\
    Conscience: Placebic & $-0.06$ & 0.1  \\
  \end{tabular}
  \caption{Linear Mixed Model results for Agreement with the AI classifier. (* = $p$ < 0.05, ** = $p$ < 0.01, *** = $p$ < 0.001). TC = Tech Comfort, Neuro= Neuroticism, Conscience= Conscientiousness}
\label{tab:lmer-agree}
\end{table}

\section{Discussion}
Increased agreement to classifications without explanation can provide insight for inclusive eXplainable AI (XAI) design, facilitating action-oriented nudges for improved user collaboration with fast-moving AI. Our findings show that individuals with higher neuroticism and low-tech comfort tend to prefer minimal or no explanations. The inclination of participants with lower comfort in technology towards preferring minimal information aligns with previous research indicating that decreased technological competence correlates with reduced reliance on technology, resulting in diminished trust in information delivered by new technological platforms~\cite{park_exploring_2019}. Furthermore, excessive information can overwhelm certain personalities and increase perceived complexity ~\cite{linde_effect_2023}, reducing efficacy for people with higher neuroticism. Moreover, perceived technological insecurity, particularly among people with high neuroticism and low-tech comfort, may contribute to a preference for minimal explanations from AI classifiers ~\cite{craig_it_2019, sundar_rise_2020, guzman_artificial_2020}. Psychologically informed approaches to AI-based systems are recommended, highlighting the importance of personalizing explanations based on user traits to improve acceptance and adaptation to AI technologies ~\cite{gentile2021evaluating}. For users with low technological comfort, providing personalized support and training resources tailored to their existing comfort levels is crucial to enhance their technology and AI comfort.

Our lack of supporting evidence for effects between conscientiousness and the preference for meaningful explanations fails to support Chiou et al.'s finding that people with high internal locus of control (LoC) are comfortable with robots exhibiting more information and control for navigation tasks~\cite{chiou_trust_2021}. LoC refers to the belief of people that their actions and behaviors affect the outcomes of events in their lives~\cite{rotter_generalized_1966}. Interestingly, participants with higher conscientiousness do not have a higher preference for meaningful explanations, that provide more control to the participants.

Designing XAI systems considering user personality and preferences is crucial for ensuring user satisfaction and XAI systems' effectiveness. Different users have varying thresholds for information complexity and detail, influenced by their personality traits, cognitive styles, and context-specific needs. User acceptance and trust in XAI systems can be enhanced by incorporating user-profiles and allowing for adjustable explanation levels. This approach respects individual user preferences and improves the overall user experience by providing information in explanations as the user requires. Such flexibility ensures that XAI systems are not just tools for transparency, but are also adaptable interfaces that enable users to engage with AI on their own terms.

\section{Conclusion}

The interplay between user personality traits, technological comfort, and explanation level presents both challenges and opportunities to create more effective and engaging XAI designs. Certain user traits like higher neuroticism and lower technological comfort, need a ``less is more'' approach to ensure accessible and inclusive XAI design. By paying attention to the diverse needs and preferences of users, XAI designers can develop systems that are more welcoming, easier to use, and conducive to human-AI collaboration.


\section{Acknowledgements}
This material is based upon work supported by the Department of Defense via Purdue University under funding agency 13000844-031 as well as Grant Thornton through an equipment grant. We would also like to thank Neelamberi Klein for helping with data analysis verification.

\balance{} 

\bibliographystyle{SIGCHI-Reference-Format}
\bibliography{sample, references, bib1}

\end{document}